\documentstyle [12pt,a4p,epsfig,amsmath,multicol]{article}
\begin{document}
\vspace*{0.6cm}

\begin{center} 
{\normalsize\bf Uniformly moving clocks in special relativity: Time dilatation, but no relativity
  of simultaneity or length contraction}
\end{center}
\vspace*{0.6cm}
\centerline{\footnotesize J.H.Field}
\baselineskip=13pt
\centerline{\footnotesize\it D\'{e}partement de Physique Nucl\'{e}aire et 
 Corpusculaire, Universit\'{e} de Gen\`{e}ve}
\baselineskip=12pt
\centerline{\footnotesize\it 24, quai Ernest-Ansermet CH-1211Gen\`{e}ve 4. }
\centerline{\footnotesize E-mail: john.field@cern.ch}
\baselineskip=13pt
\vspace*{0.9cm}
\abstract{ Time-like and space-like invariant space-time intervals are used to
 analyse measurements of spatial and temporal distances defined by two
  spatially-separated clocks in the same inertial frame. The time dilatation effect is confirmed, but not 
 `relativity of simultaneity' or `relativistic length contraction'. How these
  latter, spurious, effects arise from misuse of the Lorentz transformation is also
  explained.}
 \par \underline{PACS 03.30.+p}
\vspace*{0.9cm}
\normalsize\baselineskip=15pt
\setcounter{footnote}{0}
\renewcommand{\thefootnote}{\alph{footnote}}

    The Lorentz transformation (LT) relates space-time coordinates ($x$,$y$,$z$,$t$) of an
  event in one inertial frame, S, to those of the same event ($x'$,$y'$,$z'$,$t'$) in another
  inertial frame S'.  As is conventional, it is assumed that S' moves with uniform velocity,
  $v$, along the common $x$,$x'$ axis of the two frames. 
  \par In any actual experiment, the times $t$ and $t'$ must be measured by clocks at rest in S
   and S' respectively. Consider a clock, C1', at rest in S', observed from the frame S,
   to register time $t'$. The proper time of a clock at rest in S is denoted by $\tau$.
    The LT may be used to
   define the time-like invariant interval relation~\cite{Poincare,Mink} connecting these
   times:
   \begin{equation}
      c^2  (\Delta \tau)^2 -  (\Delta x)^2
      = c^2  (\Delta t')^2 -  (\Delta x')^2 
   \end{equation}
    where $c$ is the speed of light in vacuum, and $\Delta \tau \equiv \tau_1-\tau_2$ etc. 
    Considering two events on the
   worldline of C1', in virtue of the relations $\Delta x'= 0$ and $\Delta x = v \Delta \tau$,
    Eqn(1) gives the time dilatation (TD) formula relating $\Delta t'$, a time interval
    registered by C1', as viewed from S, to $\Delta \tau$, the 
    corresponding time interval recorded by a clock, similar to C1', but at rest in S:
    \begin{equation}
       \Delta \tau = \gamma \Delta t'
    \end{equation}
  where $ \gamma \equiv 1/\sqrt{1-(v/c)^2}$.
    Introducing such a clock at rest in S, and setting both this clock and C1' to zero,
  when they both have the same $x$-coordinate, 
     enables (2) to be written as:
     \begin{equation}
        \tau_1 = \gamma t'_1
    \end{equation}  
      where $t'_1$ is the time recorded by C1', as viewed from S, and $\tau_1$ that of the
   clock in S, as viewed in S.
   This method for synchronising
   clocks in different inertial frames has been called `system external synchronisation'
   by Mansouri and Sexl~\cite{MS}. Following Einstein~\cite{Ein1,EinSGR} it is the standard 
   way to synchronise clocks in different frames in special relativity.
   
     Carrying out exactly the same procedure for a second clock C2', lying on the $x'$-axis,
     at rest in S', at a different position to C1', gives similarly:
    \begin{equation}
        \tau_2 = \gamma t'_2
    \end{equation}
     where $t'_2$ is the time recorded by C2' as viewed fom S and $\tau_2$ that of a clock
      at rest in S, as viewed in S. 
      If now the two clocks in S that have been used to `externally synchronise'~\cite{MS}
      C1' and C2' also happen to be synchronised in S, so that $\tau_1 = \tau_2$, it follows
      from (3) and (4)
      that the clocks C1' and C2' are also synchronous, $t'_1 = t'_2$, in S', 
      ---there is no `relativity of simultaneity' (RS) effect.

    \par A space-like invariant interval between arbitary events 
    on the world lines of C1' and C2' may be defined as:
      \begin{equation}     
     (\Delta x_{12})^2 -c^2  (\Delta \tau_{12})^2 
     =  (\Delta x'_{12})^2 -c^2  (\Delta t'_{12})^2
    \end{equation}
     where $\Delta \tau_{12} \equiv \tau_1-\tau_2$ etc and the subscripts $1$ and $2$ denote the clocks
     C1' and C2' respectively.
    The spatial separation of C1' and C2' in S is, by definition, the value of $\Delta x_{12}$
     at any definite instant in S, i.e. when $\tau_1 = \tau_2 = \tau$ or $\Delta \tau_{12} = 0$. It
     follows from (3) and (4), in the case that the S-frame clocks are synchronised, that, 
     at the same instant, $t'_1 = t'_2 = t'$ or
     $\Delta t'_{12} = 0$. Thus (5) may be written, taking the positive square root of
     both sides, as:
      \begin{equation} 
     \Delta x_{12}(\Delta \tau_{12} = 0) \equiv L =  \Delta x'_{12}(\Delta t'_{12} = 0)
       \equiv L'
   \end{equation}
 
   Thus the spatial separation of the clocks is a Lorentz invariant quantity~\cite{JHF1}
    ---there is no `length contraction' (LC).

    \par As previously discussed in detail elesewhere~\cite{JHFLLT,JHFCRCS,JHFACOORD,JHFSSSC} the spurious RS and LC effects
    of conventional special relativity are the result of an incorrect use of the space-time LT
    to analyse space and time measurements. With arbitary clock synchronisation,
     the LT relating events in the frames S and S' is:
    \begin{eqnarray}
   x' & = & \gamma[x-v(\tau-\tau_0)] \\
   t'-t'_0 & = &  \gamma [\tau-\tau_0 -\frac{vx}{c^2}]
   \end{eqnarray}
    The time offsets $\tau_0$ and  $t'_0$ are specific to a particular synchronised clock and
    must be chosen in such a way as to correctly
    describe the times registered by such clocks at different spatial locations~\cite{JHFCRCS}.
     The necessity to introduce such additive constants was aleady pointed out in 
       Ref.~\cite{Ein1}, but never (to the present writer's best knowledge) implemented
      by Einstein, or any later author,
       before the work presented in Ref.~\cite{JHFLLT}.
      Placing the clocks C1' and C2' at $x'_1 = -L/2$ and $x'_2 = L/2$ respectively, in the case
     when the clocks in S are synchronised, so that $\tau_1 = \tau_2 = \tau$ and $t'_1 = t'_2 = t'$,
      the clock C1' is described by the LT:
     \begin{eqnarray}
    x'_1+\frac{L}{2} & = & \gamma[x_1+\frac{L}{2}-v\tau] = 0 \\
   t' & = &  \gamma[\tau -\frac{v(x_1+\frac{L}{2})}{c^2}]   
 \end{eqnarray} 
     and C2' by the LT:
    \begin{eqnarray}
    x'_2-\frac{L}{2} & = & \gamma[x_2-\frac{L}{2}-v\tau] = 0 \\
   t' & = &  \gamma[\tau -\frac{v(x_2-\frac{L}{2})}{c^2}]   
 \end{eqnarray}
   According to Eqns(10)-(12) C1' and C2' are externally synchronised with the synchronised
   clocks in S at the time $\tau = t' = 0 $ when  $x_1 = -L/2$ and $x_2 = L/2$. Using (9) to
 eliminate $x_1$ from (10) or (11) to eliminate $x_2$  from (12) gives:
    \begin{equation}
    \tau  = \gamma t'
    \end{equation}
    This is the time dilatation formula that is valid for all
    synchronised clocks in the frame S' when they are viewed from the frame S. The equation (13) contains no spatial
     coordinates and so is valid for synchronised clocks, one at rest in S the other at rest in S', as viewed from S, at any
      spatial locations whatever. A corollary is that simultaneity and the sign of a time interval are
      both absolute ---they are the same in all reference frames whether inertial or accelerated.
    The TD relation (13) describes the only way in 
    which the space-time geometry of special relativity differs from
   that of Galilean relativity. The equations (9) and (11),
   describing the motion of the clocks in S, are the same as the corresponding Galilean formulae.
  
  \par The spurious `RC' and `LC' effects result from the use of an incorrect LT to describe 
   a moving clock, i.e. using wrong values for the time offsets $\tau_0$ and $t'_0$ in (7) and (8).
     For example, setting $\tau = 0$  and making the substitutions $x'_1 \rightarrow x'_2$,
     $x_1 \rightarrow x_2$ and $t' \rightarrow t'_2$ in (9) and (10). That is,
       subsituting the coordinates of C2' into the LT appropriate for C1'. This gives:
   \begin{eqnarray}
    x'_2+\frac{L}{2} & = & \gamma(x_2+\frac{L}{2}) \\
   t'_2 & = &  -\gamma\frac{v(x_2+\frac{L}{2})}{c^2}  
 \end{eqnarray}
      Since setting $\tau = 0$ in (9) and (10) gives $t' \equiv t'_1 = 0$ and $x_1 = x'_1 = -L/2$,
      (14 ) and (15) may be written as:
  \begin{eqnarray}
    x'_2-x'_1 & = & \gamma(x_2 -x_1) \\
   t'_2- t'_1  & = &  -\frac{v(x'_2-x'_1)}{c^2} = -\frac{vL}{c^2} 
 \end{eqnarray}
    (16) is the LC effect ---the spatial separation of the clocks is reduced by the factor
    $1/\gamma$ in S as compared to S', and (17) is the RS effect  ---simultaneous 
    events in S: $\tau_1 = \tau_2 = 0$ are not so in S':  $t'_1 >  t'_2$ and the sign of $t'_2- t'_1$ depends 
     on that of $v$. These spurious 
   predictions result from the use of wrong values of  $\tau_0$ and $t'_0$ for the clock C2'.
   This is what has, hitherto, been universally done,
    following Einstein~\cite{Ein1,EinSGR}, when applying the LT to
    space-time measurements.
   \par The standard LT derived by Einstein is given by setting $L = 0$ in (9) and (10) or
    (11) and (12). This transformation externally synchronises clocks in S and S' at $\tau = t' = 0$
     when $x = x' = 0$. The usual application of this LT (e.g. the discussion of LC
     in Ch XII of Ref.~\cite{EinSGR}) assumes, incorrectly, that it also valid for a synchronised
      clock in S' at $x' = L$. This gives the same LC prediction as (16), and RS prediction
       as (17). The close connection between
      LC and RS in (16) and (17), discussed in~\cite{JHFRSBP}, was not mentioned by Einstein, and is typically
     not mentioned either in text books on special relativity. An exception is the book by Stephenson
    and Kilmister~\cite{SK}.
    \par The essential point made in this letter is that physics, either that underlying
     the clock mechanism, or the relativistic TD effect of Eqn(2), predicts only the {\it rate}
     of a clock, not its setting. The spurious RS and LC effects arise because
   a fixed clock {\it setting}, built into the `standard' LT ((9) and (10) with $L = 0$)
    is misinterpreted as a physical time interval. 
    \par A pendulum constitutes a clock. Its motion is completely defined by its period (determined
      by physics) and a single geometrical constant, say the angular displacement from equilibrium
       at which its kinetic energy vanishes, which is analogous to the clock offset $t'_0$. This constant
     is arbitary and not predictable.
      The RS effect of conventional special relativity is tantamount to predicting the angular
     displacement of a pendulum, in a moving frame, knowing only its length
      and the acceleration of gravity ---the essential
       physical parameters of the problem (c.f. Eqn(17), where the essential physical
     parameters are $L$, $v$ and $c$) ---an evidently nonsensical procedure.    
    \par  Experiments have recently been proposed to
    search for the existence (or not) of the RS effect~\cite{JHFRSE}.

 \par{\bf Added Note}
 \par The time intervals in the space-like invariant interval relation (5) do not correspond, as assumed in the
  derivation of (6), to the case when C1'and C2' are synchronised. Equation (5) is derived on the assumption that 
   the usual LT, given by setting $t'_0 =\tau_0 = 0$ in (7) and (8), is valid for both clocks. This implies
   that the clocks are not synchronised. Using the correct LT for each clock, (9) and (10) for
    C1' and (11) and (12) for C2', which synchronise the clocks so that $t'= \tau = 0$ when 
          $x(C1')= -L/2$ and $x(C2')= L/2$, to calculate the intervals $\Delta x'_{12}$ and $\Delta t'_{12}$ gives,
      on replacing $L$ by $L'$ on the left sides of (9) and (11):   
     \begin{eqnarray}      
    (\Delta x'_{12})^2 -c^2  (\Delta t'_{12})^2 & = & (\Delta x_{12})^2 -c^2  (\Delta \tau_{12})^2 \nonumber \\
         &   & +2 \gamma L'(\Delta x_{12}-L)-2\Delta x_{12}L-2 \gamma v\Delta \tau_{12}+L^2 +(L')^2
      \end{eqnarray}
     Setting $\Delta t'_{12} = \Delta \tau_{12} = 0$ and $\Delta x_{12} = L$ in this equation leads not to the
     relation $L = L'$ but to the identity
    \begin{equation}
    \Delta x'_{12} \equiv L' = L'
    \end{equation}
     In fact the equality of $L$ and $L'$ is established in a straightforward manner from, for example, (9)
       on replacing $L$ by $L'$ on the left side:
 \begin{equation}
 x'_1+\frac{L'}{2}  =  \gamma[x_1+\frac{L}{2}-v\tau] = 0
  \end{equation}
  The parameters $L$ and $L'$ are constants fixed by the choice of spatial  coordinate systems in S and
   S' respectively, that are independent both of time and the velocity $v$:
   \begin{eqnarray}
      x'_1(t') & = & -\frac{L'}{2}~~~~({\rm~all~}t')   \\
     x_1(\tau = 0) & = &  -\frac{L}{2}
    \end{eqnarray}
     Since $L$ and $L'$ in (20) are independent of $v$, the equation is valid for all values
    of $v$, in particular for $v = 0$, $\gamma = 1$ and $x \rightarrow x'$:
   \begin{equation}
     x'_1+\frac{L'}{2}  =  x'_1+\frac{L}{2}
   \end{equation}
     or 
   \begin{equation}
     L'  = L
   \end{equation}

  \par{\bf Acknowledgements}
  
   \par I gratefully acknowledge discussions with, or correspondence from: G.Boas,\newline B.Echenard,
  M.Gr\"{u}newald, Y.Keilman, M.Kloster, B.Rothenstein and D.Utterback, whose tenacious defence of the
   conventional 
  interpretation of the LT forced me to sharpen and condense the logical structure
  of my arguments in order to write this letter.

\end{document}